\documentclass[twocolumn,reprint,superscriptaddress,amsmath,amssymb,aps,pra]{revtex4-2}
\usepackage{paracol}
\usepackage{units}
\usepackage{graphicx,subfigure}
\usepackage{mathrsfs}
\usepackage{bm}
\usepackage{textcomp}
\usepackage{url}
\usepackage{dsfont}
\usepackage{braket}
\usepackage{xcolor}
\usepackage[compact]{titlesec}
\usepackage[colorlinks=true,linkcolor=magenta,citecolor=blue]{hyperref}%
\usepackage{todonotes}
\usepackage{empheq}
\usepackage{multirow}
\usepackage{titlesec}
\definecolor{mypink1}{rgb}{0.858, 0.188, 0.478}

\usepackage{cancel}
\titleformat{\subsubsection}
 {\normalfont\normalsize\itshape}{\thesubsubsection}{1em}{}
\titlespacing*{\subsubsection}{0pt}{3.25ex plus 1ex minus.2ex}{0ex plus.2ex}
\begin{document}
\title{Enhanced second-order sideband generation and slow-fast light via coupled opto- and magnomechanical microspheres}
\author{Abdul Wahab} 
\affiliation{Department of Physics, Jiangsu University, Zhenjiang, 212013, China}
\author{Muqaddar Abbas}
\affiliation{Ministry of Education Key Laboratory for Nonequilibrium Synthesis and Modulation of Condensed Matter, Shaanxi Province Key Laboratory of Quantum Information and Quantum Optoelectronic Devices, School of Physics,
Xi’an Jiaotong University, Xi’an 710049, China}
\author{Xiaosen Yang}
\email{yangxs@ujs.edu.cn}
\affiliation{Department of Physics, Jiangsu University, Zhenjiang, 212013, China}
\author{Yuee Xie}
\affiliation{Department of Physics, Jiangsu University, Zhenjiang, 212013, China}
\affiliation{Quantum Sensing and Agricultural Intelligence Detection Engineering Center of Jiangsu Province, Zhenjiang, Jiangsu, 212013, P. R. China}
\author{Yuanping Chen}
\email{chenyp@ujs.edu.cn}
\affiliation{Department of Physics, Jiangsu University, Zhenjiang, 212013, China}
\affiliation{Quantum Sensing and Agricultural Intelligence Detection Engineering Center of Jiangsu Province, Zhenjiang, Jiangsu, 212013, P. R. China}
\date{\today}
\begin{abstract}
In this research, we investigate second-order sideband generation (SSG) and slow-fast light using a hybrid system comprised of two coupled opto- and magnomechanical microspheres, namely a YIG sphere and a silica sphere. The YIG sphere hosts a magnon mode and a vibration mode induced by magnetostriction, whereas the silica sphere has an optical whispering gallery mode and a mechanical mode coupled via optomechanical interaction. The mechanical modes of both spheres are close in frequency and are coherently coupled by the straightway physical contact between the two microspheres. We use a perturbation approach to solve the Heisenberg-Langevin equations, offering an analytical framework for transmission rate and SSG. Using experimentally feasible settings, we demonstrate that the transmission rate and SSG are strongly dependent on the magnomechanical, optomechanical, and mechanics mechanics coupling strengths (MMCS) between the two microspheres. The numerical results show that increasing the MMCS can enhance both the transmission rate and SSG efficiency, resulting in gain within our system. Our findings, in particular, reveal that the efficiency of the SSG can be effectively controlled by cavity detuning, decay rate, and pump power. Notably, our findings suggest that modifying the system parameters can alter the group delay, thereby regulating the transition between fast and slow light propagation, and vice versa. Our protocol provides guidelines for manipulating nonlinear optical properties and controlling light propagation, with applications including optical switching, information storage, and precise measurement of weak signals.

\end{abstract}
\maketitle
\section{Introduction}\label{sec:Introduction}
Cavity magnomechanics (CMM)~\cite{doi:10.1126/sciadv.1501286,Zuo_2024}, which involves coupling a microwave (MW) cavity with a ferrimagnetic crystal such as yttrium iron garnet (YIG), has emerged as a rapidly expanding field with significant applications in contemporary quantum technologies~\cite{YUAN20221, PhysRevB.109.L041301}. This field opens new avenues for investigating the interactions between magnon, cavity, and phonon modes~\cite{PhysRevLett.129.243601,PhysRevApplied.22.044025}. Magnetic materials, particularly YIG, are noteworthy due to their high spin density, long coherence times, and strong spin-spin exchange interactions, offering a novel platform for cavity optomagnonics~\cite{Rao2019,Wolz2020,Lachance-Quirion_2019, ZHANG2023100044}. The unique dynamics of YIG enable promising applications, including long-time memory~\cite{Zhang2015, PhysRevLett.127.183202}, microwave-to-optical conversion~\cite{PhysRevB.102.064418, Chai:22}, magnon-induced nonreciprocity~\cite{PhysRevLett.123.127202, PhysRevApplied.14.014035}, enhanced tripartite interactions~\cite{PhysRevLett.130.073602}, quantum entanglement~\cite{PhysRevResearch.1.023021,10.1063/5.0015195}, and precision measurements~\cite{PhysRevLett.125.117701,doi:10.1126/science.aaz9236, PhysRevA.103.062605}, to name a few.

Simultaneously, the shape distortion of the YIG structure during magnetization induces a nonlinear interaction between the phonon and magnon modes~\cite{PhysRev.110.836}. Experimentally, strong and ultrastrong couplings between magnons and microwave photons have been achieved through magnetic dipole interaction~\cite{PhysRevLett.113.083603,Zhang2015}, providing a unique platform for exploring various quantum effects~\cite{PhysRevApplied.17.034024, PhysRevA.103.063708, PhysRevLett.128.013602, PhysRevA.99.043803,PhysRevA.108.063703, Li_2021, PhysRevA.103.053712,Xiong:24}. 
In addition, CMM, like cavity optomechanics, has been
proposed and demonstrated experimentally, where magnon-phonon interaction is introduced
through magnetostrictive force resulting in magnomechanically induced transparency (MMIT)~\cite{doi:10.1126/sciadv.1501286,PhysRevX.11.031053,PhysRevLett.129.123601,Bayati:24}. MMIT, an analog of optomechanically induced transparency~\cite{RevModPhys.86.1391}, which arises from the
interference of sidebands generated by the parametric coupling to phonons. This phenomenon further advances the study of magnetically controlled ultraslow light engineering~\cite{PhysRevApplied.15.024056,Kong:19}. Moreover, these pioneering works~\cite{doi:10.1126/sciadv.1501286,PhysRevX.11.031053,PhysRevLett.129.123601,PhysRevLett.113.083603} have also brought about a series of novel effects and applications in classical and quantum regimes~\cite{PhysRevLett.120.057202,Wang:18,PhysRevB.104.224434,WAHAB2024115436}. Notably, high-order sidebands have sparked great interest in optical communications~\cite{Lu:21}, optical and magnonic frequency combs~\cite{PhysRevLett.131.243601,PhysRevA.110.023507} and  high-sensitivity
measurement~\cite{Hodaei2017}.

The formation of high-order sidebands is essentially a nonlinear phenomenon, which can be viewed as a parametric process~\cite{PhysRevA.86.013815}. It is generally recognized that symmetrical optical sideband spectra are generated using frequency combs~\cite{doi:10.1126/science.aay3676}. As research has advanced, numerous methods for generating frequency combs have been proposed, such as optomechanical frequency combs~\cite{Miri_2018} and optomagnonic frequency combs~\cite{Liu:22}. Furthermore, frequency combs have been theoretically proposed and experimentally realized in spin waves~\cite{PhysRevLett.127.037202,10.1063/5.0090033}. The study of high-order optical sideband
generation thereby becomes an indispensable part in the field of precision measurement~\cite{Udem2002}. High-order sidebands have been previously studied in Kerr resonators~\cite{PhysRevA.97.013843}, hybrid optomechanical systems~\cite{PhysRevA.86.013815,PhysRevA.86.013815,Liu:21}, non-Hermitian systems~\cite{PhysRevA.99.033843}, and atom-cavity coupling system~\cite{PhysRevA.99.063810}. Additionally, it has been proposed that hybrid cavity magnonic systems can generate magnon-induced high-order sidebands~\cite{Xu:20,PhysRevApplied.18.044074}, which offers a novel approach for producing frequency combs in magnon spintronics~\cite{PhysRevLett.127.037202,Liu:22}, and can be utilized for precise detection of nonlinear energy spectra~\cite{PhysRevLett.128.183603} as well as magnon-based precision measurement~\cite{PhysRevApplied.13.064001}.

Nevertheless, in practice, high-order MMIT sidebands are much weaker than the probe signal, providing substantial hurdles in identifying and employing the second-order sideband~\cite{PhysRevA.109.033701,PhysRevA.107.063714}. Furthermore, in recent years, there has been tremendous success in researching fast-slow light conversion employing the cavity-magnon system~\cite{PhysRevA.102.033721,PhysRevA.108.033517}. This progress is achieved by regulating the group delay of the output light field, which is affected via rapid phase dispersion~\cite{8701447,FENG2022127781}. The slow/fast light effects in the hybrid cavity-magnon system have a wide range of applications in optical communications and interferometry~\cite{PhysRevLett.92.253201,PhysRevA.75.053807}.

These aforementioned studies motivate us to look for approaches for improving and controlling the second-order MMIT sidebands and slow/fast light in a CMM system.  
Despite having similarities to cavity optomechanics~\cite{RevModPhys.86.1391} and CMM in numerous ways, the direct analysis of second-order sideband generation (SSG) and slow/fast light using a perturbation approach in the optical domain~\cite{PhysRevA.86.013815}, as well as the influence of YIG and a silica microsphere~\cite{PhysRevLett.129.243601}, have not been well studied.

\begin{figure}
\centering
\includegraphics[width=0.48\textwidth]{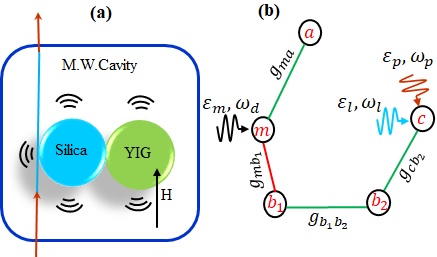} 
\caption{(a) Sketch of the hybrid system comprising a single-crystal ferromagnetic YIG sphere and a silica sphere that are in physical contact and positioned within a microwave cavity. A strong pump and a weak probe laser field drive the optical cavity (WGM), activating the optomechanical anti-Stokes scattering. The YIG (silica) sphere supports a magnon (an optical) mode and a mechanical mode coupled via the magnetostrictive (optomechanical) interaction.  The two localized mechanical modes of the two spheres are directly coupled due to their physical contact.  In addition, the magnon mode is powered by a microwave source (not shown) to boost magnomechanical coupling. (b) Diagram of the similar mode-coupling model, exhibiting the interactions between magnons, phonons, and photons.} 
\label{model}
\end{figure}

In this research, harnessing the MMIT effect, we systematically examine signal transmission, enhanced SSG, and the dynamics of slow/fast light in a hybrid system that combines opto- and magnomechanics. A YIG sphere and a silica sphere are coherently linked through direct physical contact and placed inside a microwave cavity. The YIG sphere contains two modes: a magnon mode and a mechanical vibration mode, which interact via magnetostrictive forces~\cite{Zuo_2024}. The silica sphere facilitates an optical whispering-gallery mode (WGM) and a mechanical vibration mode that link together via the optomechanical interaction~\cite{RevModPhys.86.1391}. The two mechanical modes are coupled via direct physical contact of the two spheres~\cite{PhysRevLett.129.243601}. The magnon mode is further coupled to a microwave cavity mode through magnetic dipole interaction. A strong pump and a weak probe laser field drive the optical cavity (WGM) via the optical fiber, activating the optomechanical anti-Stokes scattering. 
We derive an analytical solution for the optical transmission rate and SSG efficiency by solving the Heisenberg-Langevin equations through a perturbative approach. Based on our analytical calculations, we show that the transmission rate and SSG exhibits a
strong dependence on the magnomechanical, optomechanical, and mechanics mechanics coupling strength (MMCS) between
the two microspheres. It is found that by adjusting the MMCS, the transmission rate and SSG efficiency can be significantly enhanced, resulting in gains within our system. Furthermore, our study also shows that the efficiency of the SSG can be effectively tuned by adjusting the effective cavity detuning,  decay rate, as well as the pump power. Finally, by numerically calculating the group delay, it becomes straightforward to switch between slow and rapid light, which can be further extended by adjusting the relevant parameters. These attributes indicate that our proposed system can serve as a powerful tool for controlling light propagation, with possible applications in optical communication~\cite{PhysRevLett.92.253201}, and quantum memories~\cite{PhysRevLett.107.133601}.

The structure of this work is as follows: In~\S\ref{section: Model} we describe our suggested theoretical model and provide the steady-state solutions for our system. The findings of our research are presented in~\S\ref{section:Results}. Finally, in~\S\ref{section:Conclusions} 
we present a comprehensive conclusion that summarizes the results of our study.
In order to ensure the completeness of our work, we give detailed calculations for algebra equations and sideband parameters in~\S\ref{Appendex}. 
\section{Proposed theoretical model}\label{section: Model}
The following section describes the dynamics of our experimentally valid model, as seen in Fig~\ref{model}. Our setup comprises of a magnomechanical YIG sphere and an optomechanical silica sphere~\cite{PhysRevLett.129.243601},  
which are coherently connected by direct physical contact and placed inside a microwave cavity. The magnon mode, also known as the Kittel mode~\cite{doi:10.1126/sciadv.1501286}, is characterized by the collective motion (spin wave) of a significant number of spins in the YIG sphere. It is triggered by positioning the YIG sphere in a uniform bias magnetic field and applying a microwave drive field, such as through a loop antenna. Because of the large size of the YIG sphere, e.g., a sphere with a diameter of 200$\mu$m as in Ref~\cite{PhysRevLett.129.243601}, a dispersive type interaction is dominant between the magnon mode (at GHz) and the magnetostriction induced mechanical mode (at 10 MHz)~\cite{Zuo_2024,doi:10.1126/sciadv.1501286,PhysRevLett.121.203601,PhysRevX.11.031053,PhysRevLett.129.123601}. The magnon mode further interacts with a microwave cavity mode through magnetic dipole interaction by positioning the YIG sphere near the maximum magnetic field of the cavity mode~\cite{PhysRevLett.111.127003,PhysRevLett.113.083603,PhysRevLett.113.156401}. The silica sphere can facilitate both optical WGM and mechanical modes through radiation pressure or photoelastic effect~\cite{RevModPhys.86.1391}. The intimate contact between the two microspheres causes direct coupling of their mechanical modes. The two spheres are deliberately chosen, such that the mechanical modes are close in frequency and exhibit a linear beamsplitter-type coupling~\cite{PhysRevLett.129.243601}. 
Therefore, the Hamiltonian for such a hybrid system may be represented as ($\hbar =1$):

\begin{equation}
\begin{aligned}
H  & = \sum_{\substack{j=a, m, c \\ b_1, b_2}} \omega_j j^{\dagger} j + g_{m a}\left(m^{\dagger} a + a^{\dagger} m\right) \\
& + g_{m b_1} m^{\dagger} m \left(b_1 + b_1^{\dagger}\right) - g_{c b_2}^{\dagger} c^{\dagger} c \left(b_2 + b_2^{\dagger}\right) \\
& + g_{b_1 b_2} \left(b_1^{\dagger} b_2 + b_2^{\dagger} b_1\right) + H_{\mathrm{dri}}.
 \label{hamiltonian}
\end{aligned}
\end{equation}

Here $j= a, m, c, b_1, b_2~(j^{\dagger})$ are the annihilation (creation) operators of the microwave cavity mode, the magnon mode, the optical cavity mode, and the two mechanical modes, respectively, which satisfy the canonical commutation relation $[j,j^{\dagger}]=1$. Where $\omega_j$ represents the corresponding resonant frequencies, $g_{m a}$ signifies the cavity-magnon coupling strength, $g_{b_1 b_2}$  
is the MMCS between the two microspheres, and $g_{m b_1}$ ($g_{c b_2}$) is the bare magnomechanical (optomechanical) coupling strength, which may be substantially improved by controlling the magnon mode (optical cavity) with a powerful microwave (laser) field. The driving Hamiltonian $H_{\mathrm{dri}} / \hbar = i \varepsilon_m (m^{\dagger} e^{-i \omega_{d} t}-\text{ H.c.})+ i \sqrt{\eta_c\gamma_{c}} \varepsilon_l(c^{\dagger} e^{-i \omega_{l} t}-\text{ H.c.})+i \sqrt{\eta_c\gamma_{c}}\varepsilon_p(c^{\dagger} e^{-i \omega_{p} t}-\text { H.c.})$, pertains to the microwave (laser) fields that drive the magnon (optical cavity) modes.  The term $\varepsilon_m = \sqrt{5N/4} \gamma H_{d}$ indicates the coupling strength associated with the magnon mode and the driving magnetic field, in which~$N= \rho V$ is the total number of spins (with $V$ shows the volume, and $\rho= 4.22 \times10^{27}\textup{cm}^{-3}$ is the spin density of YIG sphare)~\cite{PhysRevLett.121.203601,PhysRevLett.124.213604}, $\gamma/2\pi$= 28 GHz/T represents the gyromagnetic ratio, with $H_{d}$ shows the field amplitude. 
The term $i \sqrt{\eta_c\gamma_{c}} \varepsilon_l(c^{\dagger} e^{-i \omega_{l} t}-\text{ H.c.})+i \sqrt{\eta_c\gamma_{c}} \varepsilon_p(c^{\dagger} e^{-i \omega_{p} t}-\text { H.c.})$ shows the interaction between the WGM and the driving laser through a fiber, whith $\gamma_{c}$ is the entire loss rate of the cavity fields, which includes the intrinsic loss rate $\gamma_{0}$ and the wave guide coupling rate $\gamma_{ex}$ ($\gamma_{c} = \gamma_{0} +\gamma_{ex}$)~\cite{PhysRevA.110.023502,PhysRevA.92.033823}. The coupling parameter~$\eta_{c} = \gamma_{ex} /\gamma_{c}$ can be frequently
adjusted, and we choose~$\eta_{c} = 0.5$ throughout our study~\cite{doi:10.1126/science.1195596}. The symbol $\varepsilon_l =\sqrt{P_{l}/\hbar \omega_l} (\varepsilon_p =\sqrt{ P_{p}/\hbar \omega_p})$ represent the amplitude of the strong pump (weak probe) field, whereas $P_{l}$ and $P_{p}$ are the powers of the corresponding pump and probe fields, respectively.

Similar to previous studies~\cite{PhysRevA.100.013813,PhysRevApplied.10.014006}, this work primarily focuses on the average response of the entire system to the probing field without considering quantum fluctuation~\cite{PhysRevA.86.013815}. To investigate the nonlinear dynamics of the system, we use the Heisenberg-Langevin equations and reduce operators to their expected values [i.e., $ o(t) = \left\langle \hat o \right\rangle (o = m,c, a, b_1, b_2)$]. So the Hamiltonian in Eq.~(\ref{hamiltonian}) leads to the following quantum Langevin equations (QLEs) with respect to $\hbar \omega_{d}\left(a^{\dagger} a+m^{\dagger} m\right)+\hbar \omega_{l} c^{\dagger} c$, can be expressed as:
\begin{equation}
\begin{aligned}
\left\langle \dot{m}\right\rangle &= -\left(i \Delta_m + \gamma_m\right) \left\langle m \right\rangle
           - i g_{m b_1}\left(\left\langle b_1 \right\rangle +\left\langle b_1^{\dagger} \right\rangle\right) \left\langle m \right\rangle \\ 
           &\quad  - i g_{m a} \left\langle a \right\rangle+ \varepsilon_m , \\
\left\langle  \dot{c} \right\rangle &= -\left(i \Delta_c + \gamma_c\right) \left\langle c \right\rangle + i g_{c b_2}(\left\langle b_2 \right\rangle +\left\langle b_2^{\dagger} \right\rangle) \left\langle c \right\rangle  \\
&\quad + \sqrt{\eta_c \gamma_c} \varepsilon_l + \sqrt{\eta_c \gamma_c} \varepsilon_p e^{-i \delta t}, \\
 \left\langle\dot{a}\right\rangle &= -\left(i \Delta_a + \gamma_a\right) \left\langle a \right\rangle - i g_{m a} \left\langle m \right\rangle, \\
\left\langle\dot{b_1}\right\rangle &= -\left(i \omega_{b_1} + \gamma_{b_1}\right) \left\langle b_1 \right\rangle - i g_{m b_1} \left\langle m^{\dagger} \right\rangle \left\langle m \right\rangle 
\\ &\quad - i g_{b_1 b_2} \left\langle b_2 \right\rangle, \\
\left\langle \dot{b_2} \right\rangle &= -\left(i \omega_{b_2} + \gamma_{b_2}\right) \left\langle {b_2} \right\rangle + i g_{c b_2} \left\langle c^{\dagger} \right\rangle \left\langle c \right\rangle - i g_{b_1 b_2} \left\langle b_1 \right\rangle,
\label{hamiltonian-1}
\end{aligned}
\end{equation}

where $\Delta_{a(m)}:=\omega_{a(m)}-\omega_{d}$, $\Delta_{c}:=\omega_{c}-\omega_{l}$, $\delta:=\omega_{p}-\omega_{l}$ are the detunings and $\gamma_j (j=a, c , m, b_1, b_2)$ is the dissipation rate of the corresponding mode.

In the particular situation of~$ \varepsilon_p\ll \varepsilon_l,  \varepsilon_m$, which adheres to the perturbative regime~\cite{PhysRevA.86.013815,PhysRevA.100.013813}, we represent the dynamical parameters as a combination of their steady-state values and small fluctuations, i.e., $o=o_s+\delta o\left(o=m, m^{\dagger}, c, c^{\dagger}, a, a^{\dagger}, b_1,  b_1^{\dagger}, b_2,  b_2^{\dagger} \right)$, where the first term indicates the steady-state values and the second term shows the small fluctuating terms. Using the perturbation expansion approach in Eq.~(\ref{hamiltonian-1}) the steady-state averages of the magnon and optical modes
are: 

\begin{equation}
\langle m_{ s}\rangle=\frac{\varepsilon_m}{\left(i \tilde{\Delta}_m+\gamma_m\right)+\frac{g_{m a}^2}{i \Delta_a+\gamma_a}}, \quad\langle c_{ s}\rangle=\frac{\varepsilon_l}{\left(i \tilde{\Delta}_c+\gamma_c\right)},
\end{equation}
with the effective detunings $\tilde{\Delta}_m= \Delta_{m}+ 2 g_{m b_1}$Re$\langle b_{1 s} \rangle$ and $\tilde{\Delta}_c= \Delta_{c}+ 2 g_{c b_2}$Re$\langle b_{2 s} \rangle$, which include the frequency shift due to the mechanical displacement jointly caused by the photo- and magnetoelastic interactions. Here we assume that the detunings i.e., $|\Delta_{a}|, \tilde{\Delta}_m, \tilde{\Delta}_c \simeq \omega_{b_1} \simeq \omega_{b_2} \gg \gamma_j (j=a, c , m, b_1, b_2)$, which leads to the following approximate expressions: $\langle m_{ s}\rangle \simeq - i \varepsilon_m /( \tilde{\Delta}_m-g_{m a}^2/\Delta_a)$ and 
$\langle c_{ s}\rangle \simeq i \varepsilon_l/   \tilde{\Delta}_{c}$. The steady-state averages of the mechanical modes are
\begin{equation}
\begin{gathered}
\left\langle b_{1 s}\right\rangle=\frac{|\langle c_{s}\rangle|^2 g_{c b_2} g_{b_1 b_2}-|\langle m_{s}\rangle|^2 g_{m b_1}\left(i \gamma_{b_2}-\omega_{b_2}\right)}{g_{b_1 b_2}^2-\left(i \gamma_{b_1}-\omega_{b_1}\right)\left(i \gamma_{b_2}-\omega_{b_2}\right)} \\
\left\langle b_{2 s}\right\rangle=\frac{|\langle c_{s}\rangle|^2 g_{c b_2}\left(i \gamma_{b_1}-\omega_{b_1}\right)-|\langle m_{s}\rangle|^2 g_{m b_1} g_{b_1 b_2}}{g_{b_1 b_2}^2-\left(i \gamma_{b_1}-\omega_{b_1}\right)\left(i \gamma_{b_2}-\omega_{b_2}\right)}.
\end{gathered}
\end{equation}

The perturbation terms of Eq.~(\ref{hamiltonian-1}) could be formed as follows:

\begin{equation}
\begin{aligned}
\dot{\delta m} &= -\left(i \tilde{\Delta}_m + \gamma_m\right) \delta m + G_{m b_1} \left(\delta b_1 + \delta b_1^{\dagger}\right) \\ 
&\quad - g_{m b_1} \left(\delta b_1 + \delta b_1^{\dagger}\right)\delta m  - i g_{m a} \delta a, \\
\dot{\delta c} &= -\left(i \tilde{\Delta}_c + \gamma_c\right) \delta c + G_{c b_2} \left(\delta b_2 + \delta b_2^{\dagger}\right) \\ 
&\quad + i  g_{c b_2} \left(\delta b_2 + \delta b_2^{\dagger}\right)\delta c + \varepsilon_p e^{-i \delta t}, \\
\dot{\delta a} &= -\left(i \Delta_a + \gamma_a\right) \delta a - i g_{m a} \delta m, \\
\dot{\delta b_1} &= -\left(i \omega_{b_1} + \gamma_{b_1}\right) \delta b_1 + G_{m b_1} \left(\delta m^{\dagger} - \delta m \right)  \\ 
&\quad - i g_{b_1 b_2} \delta b_2 - i g_{m b_1} \delta m^{\dagger} \delta m, \\
\dot{\delta b_2} &= -\left(i \omega_{b_2} + \gamma_{b_2}\right) \delta b_2 + G_{c b_2} \left(\delta c^{\dagger} - \delta c \right) \\ 
&\quad -i g_{c b_2} \delta c^{\dagger} \delta c  - i g_{b_1 b_2} \delta b_1.
\label{Heisenberg-Langevin-3} 
\end{aligned}
\end{equation}

Here, $G_{m b_1}=- i g_{m b_1} \langle m_{ s}\rangle$ and $G_{c b_2}=- i g_{c b_2} \langle c_{ s}\rangle$ shows the effective magno- and optomechanical couplings. The nonlinear terms  $g_{m b_1} \delta m (\delta b_1+ \delta b_1^{\dagger})$, $g_{c b_2} \delta c (\delta b_2+ \delta b_2^{\dagger})$, $i g_{m b_1}\delta m^{\dagger} \delta m$ and $i g_{c b_2}\delta c^{\dagger} \delta c$ are taken into consideration to generate the required second-order sidebands, whereas the higher-order sideband terms may be safely omitted owing to their small fluctuations.

To determine the amplitudes of the first (second) order sidebands, we assume that
the fluctuation terms in Eq.~(\ref{Heisenberg-Langevin-3}) have the following forms~\cite{PhysRevA.107.063714,PhysRevA.86.013815}:
\begin{multline}
\delta m =M_{1+} e^{-i \delta t}+M_{1-} e^{i \delta t} + M_{2+} e^{-2 i \delta t}+M_{2-} e^{2 i \delta t}, \\
\delta c =C_{1+} e^{-i \delta t}+C_{1-} e^{i \delta t} + C_{2+} e^{-2 i \delta t}+C_{2-} e^{2 i \delta t}, \\
\delta a_{1} =A_{1+} e^{-i \delta t}+A_{1-} e^{i \delta t} + A_{2+} e^{-2 i \delta t}+A_{2-} e^{2 i \delta t}, \\
\delta b_{1} =B_{1+} e^{-i \delta t}+B_{1-} e^{i \delta t} + B_{2+} e^{-2 i \delta t}+B_{2-} e^{2 i \delta t}, \\
\delta b_{2} =X_{1+} e^{-i \delta t}+X_{1-} e^{i \delta t} + X_{2+} e^{-2 i \delta t}+X_{2-} e^{2 i \delta t}.
\label{Heisenberg-Langevin-5}
\end{multline}

The coefficients $C_{y+}$($C_{y-}$) represent yth-order ($y=1,2,3....$) lower (upper) sidebands, respectively.  By substituting Eq.~(\ref{Heisenberg-Langevin-5}) into Eq.~(\ref{Heisenberg-Langevin-3})
and equating the coefficients of the identical order, one may compute the amplitudes of the first (second) order sidebands (the entire computations and specific constants are included in~\S\ref{Appendex}) as~\cite{PhysRevA.107.063714,PhysRevA.86.013815}.


By computing Eqs.~(\ref{Heisenberg-Langevin-7}) and~(\ref{Heisenberg-Langevin-8}) we calculate the coefficients for the first and second-order lower sidebands, revealing both the linear and nonlinear features of our system.  As a result, the coefficient for the first-order lower sideband is: 

\begin{equation}
\begin{aligned}
 C_{1+} =  \frac{\sqrt{\eta_c\gamma_{c}}\varepsilon_p}{\alpha_{11}}, \label{Heisenberg-Langevin-new}
\end{aligned}
\end{equation}
and for the second-order lower sideband is:
\begin{equation}
\begin{aligned}
C_{2+} =\frac{\Gamma_{1}}{\Psi_{17}}. \label{Heisenberg-Langevin-new-1}
\end{aligned}
\end{equation}

Considering these on hand, by employing the standard input-output relationship, i.e.,~$s_{\text {out}}=s_{\text {in}}-\sqrt{\eta_c\gamma_{c}}c (t)$~\cite{PhysRevA.31.3761},
we get the output fields of our entire system as follows~\cite{PhysRevA.86.013815,PhysRevA.98.063840}:

\begin{equation}
\begin{aligned} 
s_{\text {out}}= & s_{\text{0}} e^{-i \omega_l t}+s_{\text{1}}e^{-i \omega_p t}-\sqrt{\eta_c\gamma_{c}} C_{2-} e^{-i\left(2 \omega_p-\omega_1\right) t} \\ 
& -\sqrt{\eta_c\gamma_{c}} C_{1+} e^{-i\left(2 \omega_l-\omega_p\right) t}-\sqrt{\eta_c\gamma_{c}} C_{2+} e^{-i\left(3 \omega_1-2 \omega_p\right) t},
\end{aligned}
\end{equation}
where $s_{\text{0}}=\varepsilon_l/\sqrt{\eta_c\gamma_{c}}-\sqrt{\eta_c\gamma_{c}}c_{s}$ and $s_{\text{1}}=\varepsilon_p/\sqrt{\eta_c\gamma_{c}}-\sqrt{\eta_c\gamma_{c}}C_{1-}$. The
terms $s_{\text{0}} e^{-i \omega_l t}$ denote the output with pump frequency $\omega_l$,
while the terms~$s_{\text{1}} e^{-i \omega_p t}$ and $-\sqrt{\eta_c\gamma_{c}} C_{1+} e^{-i\left(2 \omega_l-\omega_p\right) t}$ denote the output signals corresponding to the Stokes and anti-Stokes fields, respectively. Moreover, the terms $-\sqrt{\eta_c\gamma_{c}} C_{2-} e^{-i\left(2 \omega_p-\omega_1\right) t}$ and $-\sqrt{\eta_c\gamma_{c}} C_{2+} e^{-i\left(3 \omega_1-2 \omega_p\right)}$ describe the output fields at the frequencies $\omega_{l}+2\Omega$ ($\omega_{l}-2\Omega$), corresponding to the upper (lower) SSG.

The optical transmission rate of the probe field may be calculated as:
\begin{equation}
\begin{aligned}
T = |t_{p}|^{2}= \left|1-\frac{\sqrt{\eta_c\gamma_{c}}C_{1+}}{\varepsilon_p}\right|^2.
\end{aligned}
\end{equation}
It is crucial to note that our emphasis here is on the lower SSG process. To do this, we establish the dimensionless quantity: 
\begin{equation}
\begin{aligned}
\eta_{s}= \left|-\frac{\sqrt{\eta_c\gamma_{c}}C_{2+}}{\varepsilon_p}\right|,
\label{Heisenberg-Langevin-neweff}
\end{aligned}
\end{equation}
which demonstrates the efficiency of the lower SSG.


\begin{figure*}
\includegraphics[width=1\textwidth]{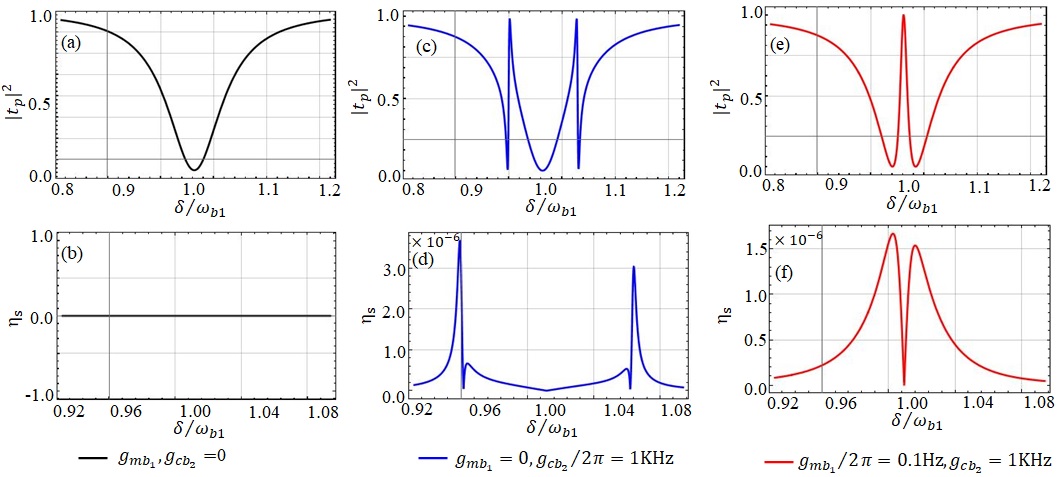}
\caption{Simulation results of the transmission rate of the probe field~$|t_{p}|^{2}$ and efficiency~($\eta_{s}$) of SSG versus optical detuning $\delta/\omega_{b_1}$. We choose for (a) and
(b)~$g_{m b_1}=0$ and $g_{c b_2}=0$, (c) and
(d)~$g_{m b_1}=0$ and $g_{c b_2}/2\pi=1$KHz, and (e) and (f)~$g_{m b_1}/2\pi=0.1$Hz and $g_{c b_2}/2\pi=1$KHz. The other parameter values are considered as: ~$\omega_{a,m}/2 \pi =7.86$~GHz,~$\omega_{b_1}/2 \pi = 20.15$~MHz,~$\omega_{b_2}/2 \pi = 20.11$~MHz,~$\gamma_{a(c),m}/2 \pi=1$ MHz,~$\lambda$=1550 nm,~$\gamma_{b_1 b_2}/2 \pi=1$ KHz, $g_{m a}/2 \pi=1$ MHz, $g_{b_1 b_2}/2 \pi=1$ MHz,~$H_{d}=3.5$ mT, $D = 200\mu$m,~$\rho= 4.22\times10^{27}\textup{cm}^{-3}$,~$\hbar = 1.054 \times 10^{-34} \ \mathrm{J \cdot s}$,~$N= 3\times10^{16}$,~$\gamma$= $2 \pi \times$ 28 GHz/T,~$P_{l} =10$~mW,~$\varepsilon_p =0.01 \varepsilon_l$,~$\Delta_a = \tilde{\Delta}_m = -\omega_{b_1}$ and~$\tilde{\Delta}_c = \omega_{b_2}$.}
\label{magnon phonon}
\end{figure*}

\section{Results}\label{section:Results}
In the preceding section, we determined the transmission rate and SSG efficiency of the output probe field for our coupled opto- and magnomechanical microsphere system. 
In this section, we present our numerical findings for investigating the transmission rate, SSG efficiency, and group delays of our system using a set of experimentally feasible parameter values~\cite{Zuo_2024,PhysRevLett.129.243601}:~$\omega_{a,m}/2 \pi =7.86$~GHz,~$\omega_{b_1}/2 \pi = 20.15$~MHz,~$\omega_{b_2}/2 \pi = 20.11$~MHz. 
We take the microwave (optical) cavity
decay rate and magnon decay
rates~$\gamma_{a(c),m}/2 \pi=1$ MHz, which are the typical values in the cavity magnonic experiments~\cite{PhysRevLett.113.083603,PhysRevLett.113.156401}, optical cavity resonance wavelength $\lambda$=1550 nm, $\gamma_{b_1,b_2}/2 \pi=1$ KHz, $g_{m a}/2 \pi=1$ MHz, $g_{b_1,b_2}/2 \pi=1$ MHz, $g_{m b_1}/2 \pi=0.1$  Hz~\cite{Zuo_2024} and the drive magnetic field $H_{d}=3.5$ mT. The parameters for the YIG sphere are as follows: diameter $D = 200\mu$m~\cite{PhysRevLett.129.243601}, the spin density~$\rho= 4.22\times10^{27}\textup{cm}^{-3}$. In addition, we take the power of the pump field~$P_{l} =10$~mW, and the amplitude of the probe filed $\varepsilon_p =0.01 \varepsilon_l$, with~$\Delta_a = \tilde{\Delta}_m = -\omega_{b_1}$ and~$\tilde{\Delta}_c = -\omega_{b_2}$. 
Unless stated otherwise, we use the aforementioned variables throughout this section.

In~\S\ref{Effect of optomechanical and magnomechanical coupling strengths on the transmission rate and SSG efficiency}, we examine the impact of magnomechanical, optomechanical, and MMCS on the transmission rate and the SSG efficiency spectrum of the output probe field. We extend the investigation to produce magnomechanically induced absorption (MMIA) and MMIT with the assistance of magnomechanical and optomechanical coupling strengths. Next, in~\S\ref{Dependence of the SSG on the detuning and decay rate}, we study the dependence of SSG on the effective cavity detuning~$\tilde{\Delta}_c$ and decay rate~$\gamma_c$. Further, in~\S\ref{Effects of the control power on the SSG generation}, we investigate the effects of pump power on the efficiency of SSG. Finally, in \S\ref{Tunable slow and fast light}, we explore the control of slow/fast light in our system.


\subsection{Effect of magnomechanical, optomechanical, and MMCS on transmission rate and SSG efficiency}\label{Effect of optomechanical and magnomechanical coupling strengths on the transmission rate and SSG efficiency}

In the subsequent subsection, we are interested in investigating the influence of opto-magnomechanical and MMCS on the transmission rate ($|t_{p}|^{2}$) and the SSG efficiency ($\eta_{s}$) of the output probe field.

We start our investigation by considering the effect of opto-magnomechanical coupling strengths on the transmission rate and the SSG efficiency. Figure~\ref{magnon phonon} displays the transmission rate ($|t_{p}|^{2}$) and the SSG efficiency ($\eta_{s}$) of the output probe field versus optical detuning~($\delta/\omega_{b_1}$) for different sets of opto-magnomechanical coupling strengths. In Fig.~\ref{magnon phonon}(a-b), we assume that both the opto-magnomechanical coupling strengths are absent ($g_{m b_1}=0$ and $g_{c b_2}=0$). Under these considerations, Fig.~\ref{magnon phonon}(a) shows a typical Lorentzian shape in the transmission spectra, with a minimum at the resonance point~($\delta=\omega_{b_1}$). This indicates that our system exhibits MMIA at~$\delta=\omega_{b_1}$. Likewise, the SSG efficiency $\eta_{s}$ spectrum, illustrated in Fig.~\ref{magnon phonon}(b), reveals a value of zero. This occurs because the nonlinear terms in Eq.(\ref{Heisenberg-Langevin-3})—specifically $g_{m b_1} \delta m (\delta b_1 + \delta b_1^{\dagger})$, $g_{c b_2} \delta c (\delta b_2 + \delta b_2^{\dagger})$, $i g_{m b_1} \delta m^{\dagger} \delta m$, and~$i g_{c b_2} \delta c^{\dagger} \delta c$—responsible for SSG, are all zero. Consequently, the SSG efficiency $\eta_{s}$ is also zero. As shown in Figs.~\ref{magnon phonon}(c-d), when the opto-magnomechanical coupling strengths are set to~$g_{m b_1}=0$ and $g_{c b_2}/2\pi=$1 KHz, we observe a central absorption dip at the resonance point~($\delta=\omega_{b_1}$). Additionally, two Fano-like resonance peaks frequently appear on both sides of the absorption window in the probe field transmission spectrum~[see Fig.~\ref{magnon phonon}(c)]. Meanwhile, the SSG spectrum also shows two asymmetric Fano peaks located at~$\delta \simeq 0.96 \omega_{b_1}$ and $\delta \simeq 1.06 \omega_{b_1}$~[see Fig.~\ref{magnon phonon}(d)]. Physically, the basic origin for the emergence of these Fano-like peaks is the presence of non-resonant interactions. For instance, in a conventional optomechanical system, if the anti-Stokes process is not resonant with the cavity frequency, the spectrum will exhibit asymmetric Fano-like profiles~\cite{PhysRevA.87.063813,PhysRevA.97.033812}. The maximum value of the SSG efficiency is about $3.4\times10^{-6}$, as shown by the solid blue curve in Fig.~\ref{magnon phonon}(d). In general, the SSG efficiency in the CMM system is fairly low, which is governed by the coupling rate and the pump power of the CMM system. The emergence of SSG is primarily due to the upconversion of the first-order sideband. In the preceding scenario~$\delta=\omega_{b_1}$, the anti-Stokes field experiences a resonant enhancement, resulting in the suppression of SSG~\cite{PhysRevA.97.013843}.

Similarly, in Figs.~\ref{magnon phonon}(e-f), when both opto-magnomechanical coupling strengths are nonzero, one may find from Fig.~\ref{magnon phonon}(e) that the 
curve of~$|t_{p}|^{2}$ exhibits a conventional MMIT profile. This profile features a transmission peak at $\delta=\omega_{b_1}$, with a linewidth of approximately 3 MHz, flanked by two deep absorption valleys at~$\delta \approx 0.985 \omega_{b_1}$ and $\delta \approx 1.015 \omega_{b_1}$. This phenomenon can be explained by destructive interference between the probe field and the anti-Stokes field, which results in the creation of a transparency window at~$\delta=\omega_{b_1}$. Correspondingly, the SSG spectrum $\eta_{s}$ shown in Fig.~\ref{magnon phonon}(f) exhibits two peaks and a local minimum at the resonance condition~$\delta = \omega_{b_1}$.

\begin{figure*}
\centering
\includegraphics[width=0.343\linewidth]{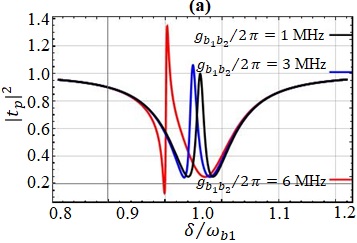}
\includegraphics[width=0.343\linewidth]{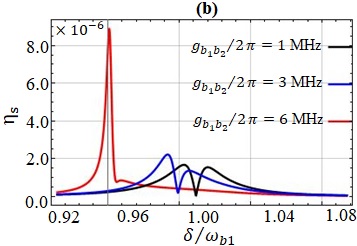}
\includegraphics[width=0.30\linewidth]{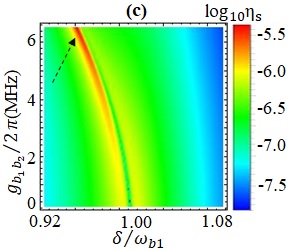}
\caption{(a-b) Transmission rate of the probe field~$|t_{p}|^{2}$ and efficiency~($\eta_{s}$) of SSG versus optical detuning $\delta/\omega_{b_1}$ is shown for different values of MMCS between the two microspheres: $g_{b_1 b_2}/2 \pi=1$ MHz, $g_{b_1 b_2}/2 \pi=3$MHz and~$g_{b_1 b_2}/2 \pi=6$ MHz. (c)~Contour map of SSG efficiency~$\eta_{s}$ (in logarithmic form) vs optical detuning $\delta/\omega_{b_1}$ and MMCS~$g_{b_1 b_2}$. All the remaining variables are similar as in Fig.~\ref{magnon phonon}(f).}
\label{Transmission a}
\end{figure*}

Let us investigate how the MMCS ($g_{b_1 b_2}$) between the two microspheres affects the optical transmission rate $|t_{p}|^{2}$ and the SSG efficiency~$\eta_{s}$ spectrum. To do this, we plotted the transmission rate~($|t_{p}|^{2}$) and the SSG efficiency ($\eta_{s}$) of the output probe field against optical detuning~($\delta/\omega_{b_1}$) for different values of the MMCS ($g_{b_1 b_2}$), as shown in Fig.~\ref{Transmission a}(a-b). In Fig.~\ref{Transmission a}(a), for a MMCS of~$g_{b_1 b_2}/2 \pi=1$ MHz, the system exhibits a transmission window at the resonance frequency ($\delta=\omega_{b_1}$), with two absorption peaks appearing symmetrically on either side of the resonance, as depicted by the black solid curve in the Fig.~\ref{Transmission a}(a). 
As anticipated, adjusting the MMCS to~$g_{b_1 b_2}/2 \pi=3$ MHz results in a higher peak amplitude of the transparency window. Additionally, the right absorption dip becomes more pronounced, while the left absorption peak narrows. Interestingly, when the MMCS is set to~$g_{b_1 b_2}/2 \pi=6$ MHz, we observe that the transparency window not only narrows and shifts away from the resonance point, but its amplitude also increases significantly, indicating amplification in our system, as indicated by the red-colored solid line in Fig.~\ref{Transmission a}(a). 

Further, the dependence of SSG efficiency~$\eta_{s}$ on the MMCS ($g_{b_1 b_2}$) is illustrated in Fig.~\ref{Transmission a}(b). It is obvious that when~$g_{b_1 b_2}/2 \pi=1$ MHz the SSG efficiency spectrum generates two peaks, with the dip at the resonance point [refer to the black solid curve in~\ref{Transmission a}(b)]. When setting $g_{b_1 b_2}/2 \pi = 3$ MHz, the right peak in the efficiency spectrum decreases, while the left peak is enhanced and the dip shifts away from the resonance point. Notably, when we set $g_{b_1 b_2}/2 \pi = 6$ MHz, the SSG efficiency spectrum mimics a line with an asymmetric peak. The right peak is more compressed and almost vanishes on the right side of the resonance point. In contrast, the left peak becomes sharper, with a peak value of~$\approx 8.9 \times 10^{-6}$ locating at~$\delta \approx 0.95 \omega_{b_1}$ [see the red solid curve in~\ref{Transmission a}(b)]. The asymmetry of the SSG spectrum is caused by constructive and destructive interference between the direct SSG process and the up-converted first-order sideband process~\cite{10.1063/1.4982167}. To provide direct insight into the influence of~$g_{b_1 b_2}$ on the efficiency of
the SSG, Fig.~\ref{Transmission a}(c) presents the contour map of the efficiency~$\eta_{s}$ (in logarithmic form) of the SSG as a function of both the MMCS and the detuning~($\delta/\omega_{b_1}$).
This figure illustrates that the efficiency of the SSG undergoes significant changes with the increase of MMCS. The maximum SSG efficiency is found in the bright red region, as indicated by the black dotted arrow [see Fig.~\ref{Transmission a}(c)].

In the above discussion, we note that increasing the MMCS ($g_{b_1 b_2}$) in our coupled opto- and magnomechanical microspheres system enhances the transmission rate and the SSG efficiency due to improved energy transfer and interaction dynamics. Stronger MMCS facilitates more efficient coupling between the mechanical modes, leading to enhanced coherence and energy exchange. This results in a more pronounced interaction between the optical and magnomechanical components,
boosting the overall transmission rate and increasing the efficiency of SSG.
\begin{figure*}
\centering
\includegraphics[width=0.343\linewidth]{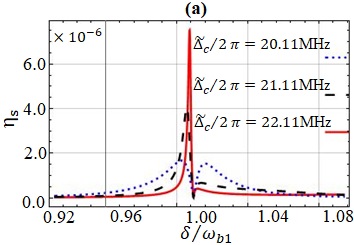}
\includegraphics[width=0.343\linewidth]{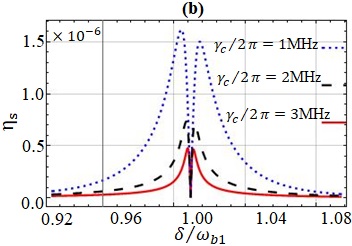}
\includegraphics[width=0.30\linewidth]{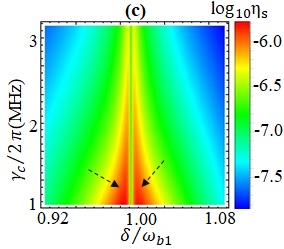}
\caption{The calculated result of SSG efficiency ($\eta_{s}$) as a function of optical detuning ($\delta/\omega_{b_1}$). (a)~For various values of effective detuning: $\tilde{\Delta}_c/2\pi = 20.11$ MHz (blue dotted curve), $\tilde{\Delta}_c/2\pi = 21.11$ MHz (black dashed curve), and $\tilde{\Delta}_c/2\pi = 22.11$ MHz (red solid curve).~(b)~For various values of decay rate:~$\gamma_{c}/2 \pi=1$ MHz (blue dotted curve),  ~$\gamma_{c}/2 \pi=2$ MHz (black dashed curve), and~$\gamma_{c}/2 \pi=3$ MHz (red solid curve). (c)~Contour map of SSG efficiency~$\eta_{s}$ (in logarithmic form) vs optical detuning $\delta/\omega_{b_1}$ and decay rate~($\gamma_{c}/2 \pi$ (MHz)). All the remaining variables are similar as in Fig.~\ref{magnon phonon}(f).}
\label{decay a}
\end{figure*}


\subsection{Dependence of the SSG
efficiency on the
detuning $\tilde{\Delta}_c$ and decay rate $\gamma_{c}$}\label{Dependence of the SSG
on the detuning and decay rate}
The effective cavity detuning~($\tilde{\Delta}_{c}$) and the decay rate~($\gamma_{c}$) also significantly impact the SSG efficiency. In the following, we demonstrate how the SSG efficiency changes with different detuning and decay rates.

In Fig.~\ref{decay a}(a), we plotted the SSG efficiency versus the scaled detuning ($\delta/\omega_{b_1}$) for various effective cavity detuning settings~($\tilde{\Delta}_{c}$). As shown in Fig.~\ref{decay a}(a), by adjusting the detuning, the SSG efficiency can be significantly modified, leading to an overall improvement in the SSG efficiency. When we set~$\tilde{\Delta}_c/2\pi = 20.11$ MHz, we observe two symmetric peaks on either side of the resonance point~($\delta=\omega_{b_1}$), with a peak value of approximately~$1.65\times10^{-6}$~[see the blue dotted line in Fig.~\ref{decay a}(a)]. Increasing the detuning to~$\tilde{\Delta}_c/2\pi = 21.11$ MHz significantly enhances the efficiency peak on the left side of the resonance point, while the peak on the right side is substantially suppressed [see black dashed line in Fig.~\ref{decay a}(a)]. Further increasing the detuning to~$\tilde{\Delta}_c/2\pi = 22.11$ MHz transforms the initially symmetric peaks into an enhanced asymmetric efficiency peak with a much narrower bandwidth and a peak value of approximately~$7.6\times10^{-6}$~[see the red solid line in Fig.~\ref{decay a}(a)]. The physical reason behind this phenomenon is that as the detuning increases, the system moves away from exact resonance, leading to an asymmetric distribution of energy. This asymmetry causes an enhancement of the SSG efficiency on one side of the resonance point while suppressing it on the other. Thus, by adjusting the effective detuning, one can control both the SSG efficiency and the symmetry of the generated peaks. This capability is essential for optimizing the performance of opto-magnomechanical devices and may provide a tool to achieve high-precision measurement~\cite{PhysRevLett.125.117701,doi:10.1126/science.aaz9236, PhysRevA.103.062605}.

In the following, we examine the impact of the decay rate ($\gamma_{c}$) on the efficiency of SSG. In Fig.~\ref{decay a}(b), the efficiency ($\eta_{s}$) of SSG is depicted as a function of the detuning ($\delta/\omega_{b_1}$) for different decay rates ($\gamma_{c}$). This figure demonstrates that as the decay rate increases~$\gamma_{c}/2 \pi=1$ MHz to~$\gamma_{c}/2 \pi=3$ MHz, the SSG  process diminishes not only around both sides of the resonance point ($\delta = \omega_{b_1}$), but also drastically reduces the linewidth of the SSG. This phenomenon can be explained as follows: Increasing the decay rate~($\gamma_{c}$) of the WGM mode reduces the SSG efficiency in a coupled opto- and magnomechanical microsphere system because a higher decay rate results in increased energy dissipation within the system. This higher dissipation reduces the effective interaction strength between the optical and magnomechanical components, lowering the efficiency of side-band production. As a result, the energy available for creating SSG is decreased, resulting in lesser efficiency~\cite{PhysRevA.99.063810}.
\begin{figure*}
\centering
\includegraphics[width=0.45\linewidth]{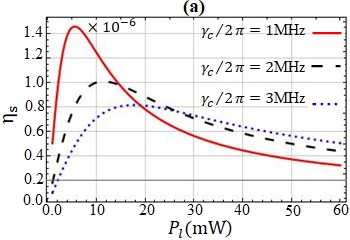}\hspace{2em}
\includegraphics[width=0.45\linewidth]{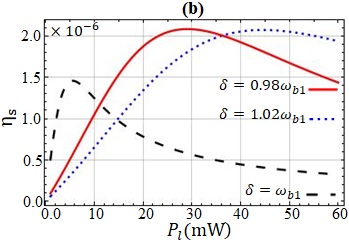}\par\medskip
\caption{The analytical results of SSG efficiency ($\eta_{s}$) as a function of pump power~$P_{l}(mW)$: (a) ~For various values of decay rate: ~$\gamma_{c}/2 \pi=1$ MHz (red solid curve),~$\gamma_{c}/2 \pi=2$ MHz (black dashed curve), and~$\gamma_{c}/2 \pi=3$ MHz (blue dotted curve) at~$\delta=\omega_{b_1}$. (b)~For various values of optical detuning :~$\delta=0.98\omega_{b_1}$~(red solid curve), $\delta=\omega_{b_1}$~(black dashed curve) and $\delta=1.02\omega_{b_1}$~(blue dotted curve).~All the remaining variables are similar as in Fig.~\ref{magnon phonon}(f).}
\label{pow plot}
\end{figure*}

To gain a more comprehensive analysis, the contour map of the calculated SSG efficiency $\eta_{s}$ (in logarithmic form) is plotted as a function of the optical detuning $\delta/\omega_{b_1}$ and the decay rate $\gamma_{c}$, as shown in Fig.~\ref{decay a}(c). The figure demonstrates that the efficiency of the SSG decreases significantly as the decay rate ($\gamma_{c}$) increases. 
The highest value of the SSG efficiency is observed at lower decay rates, as indicated by the black dotted arrows on either side of the resonance point [see Fig.~\ref{decay a}(c)]. As the decay rate increases, the color of the density plot shifts from red to blue, indicating a decrease in the SSG efficiency.

In light of the aforementioned considerations, we can describe the following aspects as influenced by the effective cavity detuning ($\tilde{\Delta}_{c}$) and decay rate ($\gamma_{c}$) on SSG efficiency ($\eta_{s}$). (i) Increasing the detuning can enhance the SSG efficiency, causing the initially symmetric peaks to merge into an asymmetric peak. Additionally, these peaks can shift and exhibit narrower linewidths based on variations in the detuning. (ii) An increase in the decay rate ($\gamma_{c}$) significantly reduces the formation of the SSG efficiency.
\subsection{Effects of pump power on the efficiency of SSG generation}\label{Effects of the control power on the SSG generation}
It is worth noting that the pump power ($P_{l}$) is a crucial parameter that will inevitably influence the efficiency ($\eta_{s}$) of SSG~\cite{PhysRevA.86.013815,PhysRevA.102.023707}. To clearly illustrate the impact of the pump power on efficiency~($\eta_{s}$) of the SSG, we plot the efficiency~($\eta_{s}$) versus
the power~($P_{l}$) in Fig.~\ref{pow plot}.


Now, let us investigate the sensitivity of SSG efficiency to variations in the decay rate ($\gamma_{c}$). According to Fig.~\ref{pow plot}(a), the enhancement of SSG is remarkable as the power is increased up to~$P_{l}= 10(mW)$, with a peak value of approximately~$1.42\times10^{-6}$~[see the red solid curve in Fig.~\ref{pow plot}(a)]. However, with a further increase in pump power, the efficiency $\eta_{s}$ sharply decreases and eventually stabilizes. The red solid, black dashed and blue dotted  lines correspond to~$\gamma_{c}/2 \pi=1$ MHz,~$\gamma_{c}/2 \pi=2$ MHz, and~$\gamma_{c}/2 \pi=3$ MHz in Fig.~\ref{pow plot}(a) respectively. It is crucial to notice that the peak efficiency value is obtained at the lowest decay rate, which is consistent with our earlier results in Fig.~\ref{decay a} (b,c). Increasing the decay rate decreases the SSG efficiency.

Correspondingly, in Fig.~\ref{pow plot}(b) we show a graph of the SSG efficiency~$\eta_{s}$ versus pump power $P_{l}(mW)$ for various optical probe field detuning~$\delta$ values. We choose the values of optical detuning~$\delta=0.98\omega_{b_1}$~(red solid curve), $\delta=\omega_{b_1}$~(black dashed curve), and $\delta=1.02\omega_{b_1}$~(blue dotted curve) [see Fig.~\ref{pow plot}(b)]. From Fig.~\ref{pow plot}(b), it is evident that for all values of probe detuning, an initial small increase in pump power leads to a sharp rise in the SSG efficiency. However, as the pump power continues to increase, the efficiency begins to decrease. It is important to note that the lowest peak efficiency is observed at the resonance point~$\delta=\omega_{b_1}$~(black dashed curve), while the highest efficiency values are achieved at  ~$\delta=0.98\omega_{b_1}$~(red solid curve) and~$\delta=1.02\omega_{b_1}$~(blue dotted curve). This observation aligns with our previous findings that maximum efficiency is attained on either side of the resonance point, whereas at the resonance point itself, the efficiency drops to zero [see Figs.~\ref{Transmission a}-\ref{decay a}]. The substantial improvement of SSG along with a much smaller bandwidth is important for prospective applications in weak signal  sensing~\cite{PhysRevA.95.033820,Xiong:17}, e.g., precise sensing of weak forces~\cite{,PhysRevLett.111.053603} and charges~\cite{PhysRevA.95.033820,Xiong:17}.
\begin{figure*}
\includegraphics[width=0.94\textwidth]{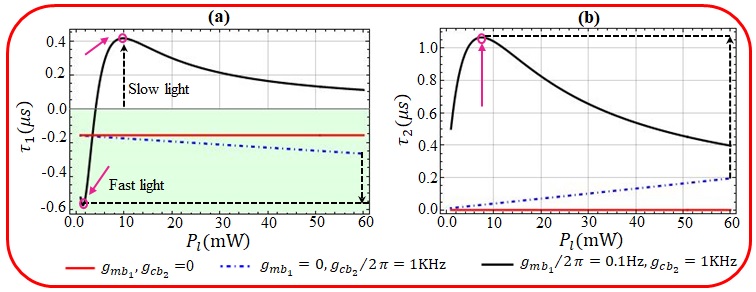}
\caption{Simulation results of the optical group delay $\tau_{1}$($\mu$s) of the probe light and the group delay $\tau_{2}$($\mu$s) of SSG as a function of the pump power~$P_{l}$ (mW), for various values of magnomechanical and optomechanical coupling strengths. The chosen parameters are: $g_{mb_1}=0$, $g_{cb_2}=0$ (red solid curve), $g_{mb_1}=0$, $g_{cb_2}/2\pi=1$ KHz (blue dotted-dashed curve), and $g_{mb_1}/2\pi=0.1$ Hz, $g_{cb_2}/2\pi=1$ KHz (black solid curve). All the remaining variables are similar as in Fig.~\ref{magnon phonon}(f), except for $\delta=\omega_{b_1}$. }
\label{group-1a}
\end{figure*}
 
\begin{figure*}
\includegraphics[width=0.94\textwidth]{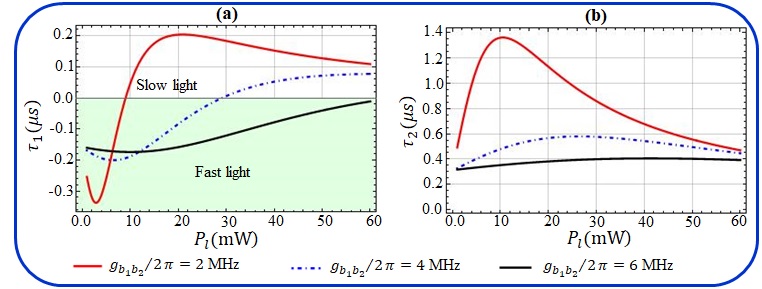}
\caption{Simulation results of the group delay~$\tau_{1}$($\mu$s) of the transmitted probe field and the group delay $\tau_{2}$($\mu$s) of SSG as a function of the pump power $P_{l}$ (mW), for several values of MMCS between the two microspheres: $g_{b_1 b_2}/2 \pi=2$ MHz~(red solid curve), $g_{b_1 b_2}/2 \pi=3$MHz~(blue dotted-dashed curve), and~$g_{b_1 b_2}/2 \pi=6$ MHz~(black solid curve). All the remaining variables are similar as in Fig.~\ref{magnon phonon}(f), except for $\delta=\omega_{b_1}$.}
\label{group-2b}
\end{figure*}
\subsection{Group delay: slow and fast light control}\label{Tunable slow and fast light}
In general, the hybrid cavity opto- and magnomechanical systems not only produce absorption and transmission effects but also enable the manipulation of light propagation, including slow/fast light phenomena~\cite{Kong:19,PhysRevA.107.063714}. The slow light phenomenon is a crucial application for Fano resonance and EIT systems~\cite{CHEN2023109242}, keeping photons inside the system for an extended period of time, which helps improve the interactions between light and matter. Generally, this feature can be quantitatively characterized by the group delay. The optical group delay of the output probe light may be described as~\cite{PhysRevApplied.18.044074,PhysRevA.100.013813}:
\begin{figure*}
\includegraphics[width=0.94\textwidth]{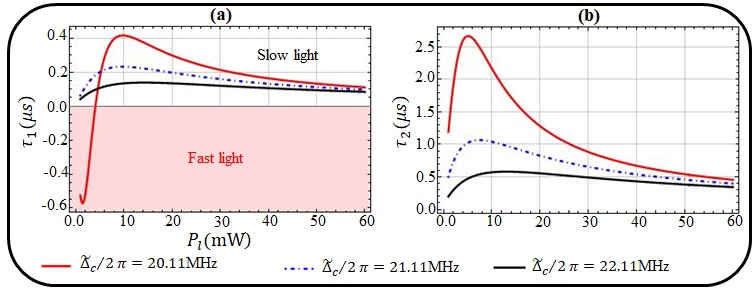}
\caption{Analytical results for the group delay~$\tau_{1}$($\mu$s) of the transmitted probe light and the group delay~$\tau_{2}$($\mu$s) of SSG as a function of the pump power $P_{l}$ (mW), for several values of effective detuning: $\tilde{\Delta}_c/2\pi = 20.11$ MHz (red solid curve), $\tilde{\Delta}_c/2\pi = 21.11$ MHz (blue dotted-dashed curve), and $\tilde{\Delta}_c/2\pi = 22.11$ MHz (black solid curve). All the remaining variables are similar as in Fig.~\ref{magnon phonon}(f), except for $\delta=\omega_{b_1}$.}
\label{group-3c}
\end{figure*}
\begin{equation}
\tau_{1}=\left.\frac{d \arg \left(t_p\right)}{d \delta}\right|_{\delta=\omega_{b_1}},
\end{equation} 
here, $\arg \left(t_p\right)$ denotes the transmission phase. The group delay of the second-order sideband may be expressed as: 
\begin{equation}
\tau_{2}=\left.\frac{d \arg \left(C_{2+}\right)}{2 d \delta}\right|_{\delta=\omega_{b_1}}.\end{equation} 

A group delay larger than zero ($\tau_{1}>0$) implies slow light, whereas a negative group delay ($\tau_{1}<0$) is associated with rapid light. Recent advances in slow-light investigations have led to developments in a variety of applications, including telecommunications and optical data storage.

A key feature of our framework is its tendency to generate either slow or rapid light by altering the system parameters. In Fig.~\ref{group-1a}, we plot the group delay $\tau_{1}$($\mu$s) of the probe light and the group delay $\tau_{2}$($\mu$s) of SSG as a function of the pump power~$P_{l}$ (mW), for various values of magnomechanical and optomechanical coupling strengths. First, we consider the scenario in which both coupling strengths are zero~($g_{m b_1}=g_{c b_2}=0$). In this case, the first-order group delay~$\tau_{1}$ remains approximately constant at -0.16$\mu$s, while the second-order group delay~$\tau_{2}$ is zero~[see red solid curve in Fig.~\ref{group-1a}(a,b)]. The physical explanation is that with zero couplings, the nonlinear terms are absent, resulting in a second-order group delay~$\tau_{2}$($\mu$s) of zero. When setting $g_{mb_1}=0$, $g_{cb_2}/2\pi=1$ KHz the first-order group delay increases continuously in the negative direction as pump power increases, indicating a fast light effect. Conversely, the second-order group delay increases steadily in the positive direction with rising pump power, reflecting a slow light effect [see blue dotted-dashed curve in Fig.~\ref{group-1a}(a,b)]. More interestingly, when both coupling strengths are present~$g_{mb_1}/2\pi=0.1$ Hz, and $g_{cb_2}/2\pi=1$ KHz, for small increases in pump power, the first-order group delay $\tau_{1}$ sharply transitions from negative to positive values (approximately 0.4$\mu$s), then decreases and eventually stabilizes. This behavior suggests that an adjustable switch from rapid to slow light can be achieved [note the black solid line in Fig.~\ref{group-1a}(a)]. In contrast, although the second-order group delay increases abruptly, it never transitions between negative and positive values, hence the total light is slow light in this situation [note the black solid line in Fig.~\ref{group-1a}(b)].

We now investigate the impact of MMCS between the two microspheres ($g_{b_1 b_2}$) on the first and second order group delays. In Fig.~\ref{group-2b}~(a,b), we illustrate the plot of first and second order group delays for various values of MMCS. Initially, when we set MMCS $g_{b_1 b_2}/2 \pi=2$ MHz for smaller pump power, both the first and second order group delays grow rapidly and then decrease [see red solid curve in Fig.~\ref{group-2b}~(a,b)]. However, only the first-order group delay transitions from negative to positive, with slow light being dominant. This indicates that both fast and slow light effects can be achieved~[see solid curve in Fig.~\ref{group-2b}~(a)]. 
In general, as the MMCS increases to $g_{b_1 b_2}/2\pi=4$ MHz, the first-order group delay transitions from fast to slow light with the rise in pump power. The peak value of the group delay significantly decreases, and the durations of both slow and fast light regions become nearly equal [see blue dotted-dashed curve in Fig.~\ref{group-2b}(a)]. Although the second-order group delay also increases with the pump power, its peak value drops from 1.38$\mu$s to 0.6$\mu$s. However, the fundamental pattern of alternating between rapid and slow light effects remains unchanged [note the blue dotted-dashed curve in Fig.~\ref{group-2b}(b)]. More interestingly, at $g_{b_1 b_2}/2\pi=6$ MHz, while the first-order group delay changes with increasing pump power, the overall light remains fast [as seen by the black solid curve in~\ref{group-2b}(a)]. The fundamental physical reason is that by adding active gain to the system, rapid light may be observed in experiments~\cite{PhysRevA.92.023855,Wang2000}. This active gain is clearly visible in our previous results [as seen by the red solid curve in Fig.~\ref{Transmission a}(a)]. It is distinctly observable that the performance of the second-order sideband group delay changes slightly at $g_{b_1 b_2}/2\pi=6$ MHz [see black solid curve in Fig.~\ref{group-2b}(b)].

Finally, we study how to control group delay via effective detuning $\tilde{\Delta}_c$. In Fig.~\ref{group-3c}, we display the first and second-order group delays as a function of the pump power $P_{l}$ and different values of effective detuning. In Fig.~\ref{group-3c}(a), when we choose~$\tilde{\Delta}_c/2\pi = 20.11$ MHz [see red solid curve in Fig.~\ref{group-3c}(a)] the group delay experiences the conversion from fast to slow light as the pump power increases. 
The delay reaches a peak value, then decreases, and eventually stabilizes at a saturated numerical value with further increases in pump power. However, the second-order group delay increases rapidly, then decreases, and ultimately stabilizes as the power increases, without undergoing any transitions [see red solid curve in Fig.~\ref{group-3c}(b)]. On increasing effective detuning~$\tilde{\Delta}_c/2\pi = 21.11$ to~$\tilde{\Delta}_c/2\pi = 22.11$ 
in both the first and second-order group delay cases, although the peak value of the group delay decreases, slow light remains dominant over most of the region [see blue  dotted-dashed and black solid curves in Fig.~\ref{group-3c}(a,b)]. Consequently, with regulating the effective detuning the slow-to-fast light can be easily achieved in our system. These findings suggest that our system can act as a tunable switch that can be adjusted using a variety of system parameters. This capability can be leveraged in optical storage or quantum communication applications~\cite{PhysRevA.91.043843,Li:17,PhysRevLett.107.133601}.
\section{Conclusions}\label{section:Conclusions}
In conclusion, we have presented a practical and efficient scheme to enhance and steer the MMIT, SSG, and group delay employing two coupled opto- and magnomechanical microspheres placed
inside a microwave cavity. We obtained analytical formulations for the transmission rate and SSG by solving the Heisenberg-Langevin equations using a perturbation approach. We have shown that the transmission rate and SSG exhibits a strong dependence on the magnomechanical, optomechanical, and MMCS between the two microspheres. It was demonstrated that modulating MMCS can boost the transmission rate and SSG efficiency while simultaneously transforming two symmetrical SSG peaks into an asymmetric Fano-like peak. In particular, we have analyzed the effects of effective cavity detuning, decay rate, and pump power on SSG efficiency. 
Finally, by examining the group delay of the probe light, we investigated the conditions for slow/fast light propagation in our system, which may be adjusted by modifying various system variables. Beyond its fundamental scientific significance, our research has important practical implications. These findings are applicable to real-world experiments where controlling light transmission is crucial, potentially benefiting applications such as microwave-optical conversion modulation, optical storage, and high-precision measurement. 
\section*{Acknowledgements}\label{section:Acknowledgements}
Abdul Wahab is grateful for the financial assistance granted by the China Postdoctoral Research Council and Jiangsu University. This project was funded by the Natural Science Foundation of Jiangsu Province (Grant No. BK20231320), National Natural Science Foundation of China (Grant No. 12174157) and Jiangsu Funding Program for Excellent Postdoctoral Talent No.2024ZB867. 
\appendix
\section{Appendex}\label{Appendex}
\subsection{Calculations of first-order sideband}
In this appendix, we present detailed calculations to obtain the amplitudes of the first-order sideband:
\begin{equation}
\begin{aligned}
M_{1+}{h_1} &= G_{m b_1} (B_{1-}^{*}+ B_{1+}) - i g_{m a} A_{1+}, \\
M_{1-}{h_2} &= G_{m b_1} (B_{1+}^{*}+ B_{1-}) - i g_{m a} A_{1-}, \\
C_{1+}{h_3} &= G_{c b_2} (X_{1-}^{*}+ X_{1+}) + \varepsilon_p, \\
C_{1-}{h_4} &= G_{c b_2} (X_{1+}^{*}+ X_{1-}), \\
A_{1+}{h_5} &= - i g_{m a} M_{1+}, \\
A_{1-}{h_6} &= - i g_{m a} M_{1-}, \\
B_{1+}{h_7} &= G_{m b_1} (M_{1-}^{*} - M_{1+}) - i g_{b_1 b_2} X_{1+}, \\
B_{1-}{h_8} &= G_{m b_1} (M_{1+}^{*} - M_{1-}) - i g_{b_1 b_2} X_{1-}, \\
X_{1+}{h_9} &= G_{c b_2} (C_{1-}^{*} - C_{1+}) - i g_{b_1 b_2} B_{1+}, \\
X_{1-}{h_{10}} &= G_{c b_2} (C_{1+}^{*} - C_{1-}) - i g_{b_1 b_2} B_{1-}.\label{Heisenberg-Langevin-7}
\end{aligned}
\end{equation}
with $h_1 = -i \delta + i \tilde\Delta_{m} +  \gamma_m$, $h_2 = i \delta + i \tilde\Delta_{m} +  \gamma_m$, $h_3 =-i \delta + i \tilde\Delta_{c} +  \gamma_c$, $h_4 = i \delta + i \tilde\Delta_{c} +  \gamma_c$,  $h_5 = - i \delta+ i \Delta_{a}+\gamma_a$, $h_6 = i \delta+ i \Delta_{a}+\gamma_a$, $h_7 = -i \delta + i \omega_{b_1} + \gamma_{b_1}$, $h_8 = i \delta + i \omega_{b_1} + \gamma_{b_1}$,~$h_9 = -i \delta + i \omega_{b_2} + \gamma_{b_2}$, and $h_{10} = i \delta + i \omega_{b_2} + \gamma_{b_2}$.

After solving Eq.~(\ref{Heisenberg-Langevin-7}) for ($C_{1+}$) we get the following constants for first-order side-band:
\begin{align*}
\alpha_1 = &\frac{\left[(-h_1 + h_2^{*}) h_5 h_6^{*} + (h_5 - h_6^{*}) g_{m a}^2 \right] (h_4^{*} h_9 - 2 G_{c b_2}^2) G_{m b_1}^2} 
{(h_1 h_5 + g_{m a}^2) (h_2^{*} h_6^{*} + g_{m a}^2)},\nonumber \\
\alpha_2 = &\frac{\left[(h_1 -h_2^{*}) h_5 h_6^{*} + (-h_5 + h_6^{*}) g_{m a}^2 \right] (h_{10}^{*} h_4^{*} + 2 G_{cb2}^2) G_{m b_1}^2}
{(h_1 h_5 + g_{ma}^2) (h_2^{*} h_6^{*} + g_{ma}^2)},\nonumber \\
\alpha_3 = &\left[ h_7 + \left( \frac{2 h_5}{h_1 h_5 + g_{ma}^2} - \frac{2 h_6^{*}}{h_2^{*} h_6^{*} + g_{ma}^2} \right) G_{mb1}^2 \right],\nonumber \\
\alpha_4 = &\ i h_4^{*} g_{b_1 b_2} \left( h_4^{*} g_{b_1 b_2}^2 + h_7 (h_4^{*} h_9 - G_{c b_2}^2) + \alpha_1 \right),\nonumber \\
\alpha_5 = &\ i h_4^{*} g_{b_1 b_2} \left(h_7 G_{c b_2}^2 - \alpha_2 \right),\nonumber \\
\alpha_6 = &-i h_4^{*} g_{b_1 b_2} \big(-h_{10}^{*} h_4^{*} h_8^{*} \alpha_4 + h_4^{*} h_7 h_9 \alpha_5 \\
&\quad + h_4^{*} (-\alpha_4 + \alpha_5) g_{b_1 b_2}^2  - (h_7 + h_8^{*}) (\alpha_4 + \alpha_5) G_{c b_2}^2 \big), \nonumber \\
\alpha_7 = &\ h_4^{*2} g_{b_1 b_2} G_{c b_2} \bigg(i (h_7 + h_8^{*}) \alpha_5 + h_4^{*} g_{b_1 b_2} \big(h_{10}^{*}h_4^{*} h_8^{*}  \\
&\quad + h_4^{*} g_{b_1 b_2}^2 + (h_7 + h_8^{*}) G_{c b_2}^2\big) \alpha_3\bigg),\nonumber \\
\alpha_8 = &\ (i h_4^{*} g_{b_1 b_2} \big( h_{10}^{*}h_4^{*} h_8^{*} + h_4^{*} g_{b_1 b_2}^2 + (h_7 + h_8^{*})  G_{c b_2}^2 \big)) \alpha_6 ,\nonumber \\
\alpha_9 = &\ (i h_4^{*2} (h_7 + h_8^{*}) g_{b_1 b_2}G_{c b_2}) \alpha_6 \\
&+ (i h_4^{*} g_{b_1 b_2} (h_4^{*} h_7 h_9 + h_4^{*} g_{b_1 b_2}^2 - (h_7 + h_8^{*}) G_{c b_2}^2)) \alpha_7 ,\nonumber \\ 
\alpha_{10} &=\left( \frac{\alpha_{7}}{\alpha_{6}} + \frac{\alpha_{9}}{\alpha_{8}}\right), \nonumber \\
\alpha_{11} &= h_3 - G_{c b_2} \alpha_{10}. \nonumber
\end{align*}

\subsection{Calculations of second-order side-band}\label{Appendex-1}
The second subgroup exhibits non-linear response (second-order side-band) and  described as:
\begin{equation}
\begin{aligned}
M_{2-}{l_2} &= i g_{m a}A_{2-} + i g_{m b_1}(M_{1-})(B_{1-}+ B_{1+}^{*}) \\
            &\quad - G_{m b_1} (B_{2+}^{*}+ B_{2-}), \\
C_{2+}{l_3} &= G_{c b_2} (X_{2-}^{*}+ X_{2+}) + i g_{c b_2}(C_{1+})(X_{1+}+X_{1-}^{*}), \\
C_{2-}{l_4} &= G_{c b_2} (X_{2+}^{*}+ X_{2-}) + i g_{c b_2}(C_{1-})(X_{1-}+X_{1+}^{*}), \\
A_{2+}{l_5} &= - i g_{m a} M_{2+}, \\
A_{2-}{l_6} &= - i g_{m a} M_{2-}, \\
B_{2+}{l_7} &= i g_{b_1 b_2}X_{2+} + i g_{m b_1}(M_{1+}M_{1-}^{*}) \\
            &\quad - G_{m b_1} (M_{2-}^{*} - M_{2+}), \\
B_{2-}{l_8} &= i g_{b_1 b_2}X_{2-} + i g_{m b_1}(M_{1-}M_{1+}^{*}) \\
            &\quad - G_{m b_1} (M_{2+}^{*} - M_{2-}), \\
X_{2+}{l_9} &= G_{c b_2} (C_{2-}^{*} - C_{2+}) - i g_{b_1 b_2} B_{2+} \\
            &\quad + i g_{c b_2}(C_{1-}^{*} C_{1+}), \\
X_{2-}{l_{10}} &= G_{c b_2} (C_{2+}^{*} - C_{2-}) - i g_{b_1 b_2} B_{2-} \\
              &\quad - i g_{c b_2}(C_{1+}^{*} C_{1-}). \label{Heisenberg-Langevin-8}
\end{aligned}
\end{equation}\\
 with $l_1 = 2 i \delta - i \tilde\Delta_{m} -  \gamma_m$, $l_2 = -2 i \delta - i \tilde\Delta_{m} -  \gamma_m$, $l_3 =-2i \delta + i \tilde\Delta_{c} +  \gamma_c$, $l_4 =2 i \delta + i \tilde\Delta_{c} +  \gamma_c$,  $l_5 = -2 i \delta+ i \Delta_{a}+\gamma_a$, $l_6 =2 i \delta+ i \Delta_{a}+\gamma_a$, $l_7 = 2i \delta - i \omega_{b_1} - \gamma_{b_1}$, $l_8 =-2 i \delta - i \omega_{b_1} - \gamma_{b_1}$,~$l_9 = -2i \delta + i \omega_{b_2} + \gamma_{b_2}$, and $l_{10} =2 i \delta + i \omega_{b_2} + \gamma_{b_2}$.\\
 After solving Eq.~(\ref{Heisenberg-Langevin-8}) for ($C_{2+}$) we get the following constants for  second-order side-band:
\begin{align*}
~~~~~~~~~~~~~~~~~~~~~~~~~~~~~~~~~~~~~~~~~\Psi_1 = &\ 2i g_{c b_2} G_{c b_2} G_{m b_1}^2 \Bigg( \frac{l_5}{l_1 l_5 - g_{m a}^2} + \frac{l_6^*}{-g_{m a}^2 + l_2^* l_6^*} \Bigg), \nonumber \\
\Psi_2 = &\ 2i g_{c b_2} G_{m b_1}^2 l_4^* \Bigg( \frac{l_5}{l_1 l_5 - g_{ma}^2} + \frac{l_6^*}{-g_{ma}^2 + l_2^* l_6^*} \Bigg), \nonumber \\
\Psi_3 = &\ \frac{G_{m b_1}^2 \left(-2 G_{c b_2}^2 + l_{10}^* l_4^* \right) \left(l_5 \left( l_1 + l_2^* \right) l_6^* - g_{ma}^2 \left( l_5 + l_6^* \right) \right)}{ \left( l_1 l_5 - g_{ma}^2 \right) \left( g_{ma}^2 - l_2^* l_6^* \right)}, \nonumber \\
\Psi_4 = &\ \frac{G_{m b_1}^2 \left( 2 G_{c b_2}^2 + l_9 l_4^* \right) \left( l_5 \left( l_1 + l_2^* \right) l_6^* - g_{ma}^2 \left( l_5 + l_6^* \right) \right)}{ \left( l_1 l_5 - g_{ma}^2 \right) \left( g_{ma}^2 - l_2^* l_6^* \right) }, \nonumber \\ 
\Psi_5 = & \frac{ l_7 G_{c b_2}^2 + (\Psi_3) }{-i g_{b_1 b_2} \left(-2 l_7 G_{c b_2}^2 + 2 g_{b_1 b_2}^2 l_4^* - l_4^* \left( G_{c b_2}^2 + l_{10}^* l_4^* \right) l_8^* \right)}, \nonumber \\
\Psi_6 = & -l_7 G_{c b_2}^2 + \left( l_7 l_9 - g_{b_1 b_2}^2 \right) l_4^* + (\Psi_4) \\
& + i \Psi_5 g_{b_1 b_2} \left( \left(-l_7 l_9 + g_{b_1 b_2}^2 \right) l_4^* + G_{c b_2}^2 \left( l_7 + l_4^* l_8^* \right) \right), \nonumber \\
\Psi_7 = & G_{c b_2} L_4^* \left(-l_7 + 2 G_{m b_1}^2 \left( \frac{l_5}{l_1 l_5 - g_{ma}^2} + \frac{l_6^*}{- g_{ma}^2 + l_2^* l_6^*} \right) \right) \\
& + i \Psi_5 g_{b_1 b_2} \left( l_7 + l_4^* l_8^* \right), \nonumber \\
\Psi_8 =  & -\Psi_1 - \Psi_5 g_{b_1 b_2} g_{c b_2} G_{c b_2} (l_7 + l_4^* l_8^*), \nonumber \\
\Psi_9 = & \Psi_2 - g_{c b_2} l_4^* \left( -i l_7 + \Psi_5 g_{b_1 b_2} (l_7 + l_4^* l_8^*) \right), \nonumber \\
\Psi_{10} = & \frac{i g_{b_1 b_2} \left( (-l_7 l_9 + g_{b_1 b_2}^2) l_4^* + G_{c b_2}^2 (l_7 + l_4^* l_8^*) \right)}{\Psi_6}, \nonumber \\
\Psi_{11} =& -\Psi_7 \Psi_{10} + i g_{b_1 b_2} G_{c b_2} l_4^* (l_7 + l_4^* l_8^*), \nonumber \\
\Psi_{12} =& -\Psi_8 \Psi_{10} - g_{b_1 b_2} g_{c b_2} G_{c b_2} (l_7 + l_4^* l_8^*)
, \nonumber \\
\Psi_{13} =& \Psi_8 \Psi_{10} + g_{b_1 b_2} g_{c b_2} G_{c b_2} (l_7 + l_4^* l_8^*)
, \nonumber \\
\Psi_{14} =& \Psi_9 \Psi_{10} - g_{b_1 b_2} g_{c b_2} l_4^* (l_7 + l_4^* l_8^*), \nonumber \\
\Psi_{15} =& \frac{i G_{c b_2}}{g_{b_1 b_2} \left( -2 l_7 G_{c b_2}^2 + 2 g_{b_1 b_2}^2 l_4^* - l_4^* \left(G_{c b_2}^2 + l_{10}^* l_4^* \right) l_8^* \right)}, \nonumber \\
\Psi_{16} =& \frac{ G_{c b_2}}{\Psi_{6}}, \nonumber \\
\Psi_{17} =& l_3 - \Psi_{15} \Psi_{11} - \Psi_{16} \Psi_{7}, \nonumber \\
\Psi_{18} =& \Psi_{12} \Psi_{15} + \Psi_{16} \Psi_{8}, \nonumber \\
\Psi_{19} =& \Psi_{15} \Psi_{13} - \Psi_{16} \Psi_{8}, \nonumber \\
\Psi_{20} =&  \Psi_{15} \Psi_{14} + \Psi_{16} \Psi_{9}, \nonumber \\
\Psi_{21} =& i l_7 \left( \Psi_{10} \Psi_{15} - \Psi_{16} \right) g_{c b_2}, \nonumber \\
\Psi_{22} =& i \left( \Psi_{10} \Psi_{15} - \Psi_{16} \right) g_{b_1 b_2} g_{m b_1} l_4^*, \nonumber \\
\end{align*}

\begin{widetext}
\[
\Psi_{23} = \frac{h_5 h_6^{*} \Psi_{22} \left( h_4 \left( -h_9 \alpha_7 \alpha_8 + h_{10}^* \alpha_6 \alpha_9 \right) - 2 h_4 \alpha_6 \alpha_8 G_{c b_2} + 2 \left( \alpha_7 \alpha_8 + \alpha_6 \alpha_9 \right) G_{c b_2}^2 \right)^2 G_{m b_1}^2}{\left( g_{b_1 b_2}^2 \left( h_1 h_5 + g_{ma}^2 \right) \left( h_2^{*} h_6^{*} + g_{ma}^2 \right) \right)}, 
\]
\[
\Gamma_1 = \frac{\varepsilon_p^2}{h_4^{*2} \alpha_{11}^2 \alpha_6^2 \alpha_8^2} \left[ h_4 \left( \alpha_7 \alpha_8 + \alpha_6 \alpha_9 \right) \left( i h_4 \alpha_6 \alpha_8 g_{cb2} + \left( -\alpha_6 \alpha_8 \Psi_{20} + \alpha_6 \alpha_9 \left( \Psi_{18} + \Psi_{21} \right) + \alpha_7 \alpha_8 \left( -\Psi_{19} + \Psi_{21} \right) \right) G_{c b_2} \right) + \Psi_{23} \right].
\]
\end{widetext}
%

\end{document}